\documentclass[nature, preprint,superscriptaddress,showpacs,amsmath, amssymb,floatfix]{revtex4-1}
\usepackage{graphicx, nicefrac}
\usepackage[thinspace,thinqspace,amssymb]{SIunits}
\usepackage{color, colortbl, hyperref}
	\definecolor{LinkColor}{rgb}{0.45,0,0}
	\definecolor{UrlColor}{rgb}{0,0,0.45}
	\definecolor{CiteColor}{rgb}{0,0.45,0}
	\hypersetup{%
	linkcolor=LinkColor,%
	filecolor=LinkColor,%
	menucolor=LinkColor,%
	citecolor=CiteColor,%
	urlcolor=UrlColor,%
	colorlinks=true%
	}

\newcommand{\cooltheta}{_{\,_{\!\!\Large\theta}}}

\begin{document}
\title{Filming the formation and fluctuation of Skyrmion domains by cryo-Lorentz Transmission Electron Microscopy}

\author{J.~Rajeswari\footnote{\label{note1}These authors contributed equally to this work.}}
\affiliation{Laboratory for Ultrafast Microscopy and Electron Scattering, ICMP, \'{E}cole Polytechnique F\'{e}d\'{e}rale de Lausanne, CH-1015, Lausanne, Switzerland.}

\author{H.~Ping\textsuperscript{\ref{note1}}}
\footnotetext{footnote with two references}
\affiliation{Laboratory for Quantum Magnetism, ICMP, \'{E}cole Polytechnique F\'{e}d\'{e}rale de Lausanne, CH-1015, Lausanne, Switzerland.}

\author{G.~F.~Mancini}
\affiliation{Laboratory for Ultrafast Microscopy and Electron Scattering, ICMP, \'{E}cole Polytechnique F\'{e}d\'{e}rale de Lausanne, CH-1015, Lausanne, Switzerland.}

\author{Y.~Murooka}
\affiliation{Laboratory for Ultrafast Microscopy and Electron Scattering, ICMP, \'{E}cole Polytechnique F\'{e}d\'{e}rale de Lausanne, CH-1015, Lausanne, Switzerland.}

\author{T.~Latychevskaia}
\affiliation{Physics Department, University of Zurich, Winterthurerstrasse 190, 8057 Zurich, Switzerland}

\author{D.~McGrouther}
\affiliation{SUPA School of Physics and Astronomy, University of Glasgow, G12 8QQ, United Kingdom.}

\author{M.~Cantoni}
\affiliation{Centre Interdisciplinaire de Microscopie Electronique,\'{E}cole Polytechnique F\'{e}d\'{e}rale de Lausanne, CH-1015, Lausanne, Switzerland.}

\author{E.~Baldini}
\affiliation{Laboratory for Ultrafast Microscopy and Electron Scattering, ICMP, \'{E}cole Polytechnique F\'{e}d\'{e}rale de Lausanne, CH-1015, Lausanne, Switzerland.}

\author{J.~S.~White}
\affiliation{Laboratory for Neutron Scattering and Imaging, Paul Scherrer Institut, CH-5232 Villigen, Switzerland.}

\author{A.~Magrez}
\affiliation{Competence in Research of Electronically Advanced Materials,\'{E}cole Polytechnique F\'{e}d\'{e}rale de Lausanne, CH-1015, Lausanne, Switzerland.}

\author{T.~Giamarchi}
\affiliation{Department of Quantum Matter Physics, University of Geneva, 24 Quai Ernest-Ansermet, CH-1211 Geneva, Switzerland.}

\author{H.~M.~R\o nnow}
\affiliation{Laboratory for Quantum Magnetism, ICMP, \'{E}cole Polytechnique F\'{e}d\'{e}rale de Lausanne, CH-1015, Lausanne, Switzerland.}

\author{F.~Carbone}
\email{fabrizio.carbone@epfl.ch}
\affiliation{Laboratory for Ultrafast Microscopy and Electron Scattering, ICMP, \'{E}cole Polytechnique F\'{e}d\'{e}rale de Lausanne, CH-1015, Lausanne, Switzerland.}


\begin{abstract}

Magnetic skyrmions are promising candidates as information carriers in logic or storage devices thanks to their robustness, guaranteed by the topological protection, and their nanometric size. Currently, little is known about the influence of parameters such as  disorder, defects or external stimuli, on the long-range spatial distribution and temporal evolution of the skyrmion lattice. Here, using a large ($7.3\times 7.3 \mu \text{m}^2$) single crystal nano-slice of Cu$_2$OSeO$_3$, we image up to 70,000 skyrmions, by means of cryo-Lorentz Transmission Electron Microscopy as a function of the applied magnetic field. The emergence of the skyrmion lattice from the helimagnetic phase is monitored, revealing the existence of a glassy skyrmion phase at the phase transition field, where patches of an octagonally distorted skyrmion lattice are also discovered. In the skyrmion phase, dislocations are shown to cause the emergence and switching between domains with different lattice orientations and the temporal fluctuations of these domains is filmed. These results demonstrate the importance of direct-space and real-time imaging of  skyrmion domains for addressing both their long-range topology and stability.

\end{abstract}


\maketitle
\section*{Introduction}

In a non-centrosymmetric chiral lattice, the competition between the symmetric ferromagnetic exchange, the anti-symmetric Dzyaloshinskii-Moriya interaction (DMI) and an applied magnetic field can stabilize a highly ordered spin-texture presenting as a  hexagonal lattice of spin-vortices called skyrmions  \cite{PhysRevLett.87.037203, nagaosa2013topological, fert2013skyrmions, do2009skyrmions}.

Magnetic skyrmions have been experimentally detected in materials having the B20-type crystal structure such as MnSi \cite{Mühlbauer13022009}, Fe$_{1-x}$Co$_x$Si \cite{munzer2010skyrmion, Milde31052013}, FeGe \cite{yu2011near}, Cu$_2$OSeO$_3$ \cite{seki2012observation} and recently also on systems like GaV$_4$S$_8$ \cite{2015arXiv150208049K} and beta-Mn-type alloys \cite{2015arXiv150305651T}. Small-angle neutron scattering (SANS) studies of bulk solids evidenced the formation of an hexagonal skyrmion lattice confined in a very narrow region of temperature and magnetic field (T-B) in the phase diagram \cite{Mühlbauer13022009,munzer2010skyrmion}. In thin films and thinly cut slices of the same compounds instead, skyrmions can be stabilized over a wider T-B range as revealed by experiments using cryo-Lorentz transmission electron microscopy (LTEM) \cite{tonomura2012real,yu2010real}. Furthermore, it was proposed that skyrmions can also exist as isolated objects before the formation of the ordered skyrmion lattice in the proximity of the phase transition \cite{sampaio2013nucleation,yu2010real}. A recent resonant X-ray diffraction experiment also suggested the formation of two skyrmion sublattices giving rise to regular superstructures \cite{langner2014coupled}. 

On reducing the geometry of interest to two-dimensions (2D), long-range ordering can be significantly altered by the presence of defects and  disorder. Indeed, the competition between order and disorder within the context of lattice formation continues to be an issue of fundamental importance. Here Condensed Matter systems are well-known to provide important test-beds for exploring theories of structural order in solids and glasses. An archetypal, and conceptually relevant example is the superconducting vortex lattice, where real-space imaging studies allow direct access to the positional correlations and local coordination numbers \cite{PhysRevB.81.014513, guillamon2014enhancement, zehetmayer2015vortex}. Up until now, however, analogous studies of skyrmion lattices have not been reported even though (as for superconducting vortices) it is well-known that defects and dislocations present in a sample can pin the motion of skyrmions induced by external perturbations such as an electric field \cite{PhysRevLett.113.107203} or a magnetic field \cite{langner2014coupled}. This competition between disorder and elasticity will clearly give rise to a complex energy landscape promoting diverse metastable states \cite{klein_nature}, and superstructures \cite{PhysRevB.87.214419, OlsonReichhardt201452}. Furthermore, up to now, previous imaging studies of skyrmion lattices could probe only the short range order due to limitations in the size of the imaged area and its homogeneity. 

In this paper, by systematic observations using cryo-LTEM, we reveal the magnetic field-dependent evolution of the skyrmion-related spin textures in Cu$_2$OSeO$_3$ thin plate and study their long-range ordering properties imaging up to $\approx{1000}$ lattice constants. The different phases of the spin textures are analyzed with  state-of-the-art methods to unravel their spatial properties. At low magnetic fields, the coexistence of two helical domains is observed, in contrast to previous studies \cite{seki2012observation}; the angle between the two helices' axis is retrieved \textit{via} a reciprocal space analysis. At the magnetic field close to the helical-skyrmion phase transition, evidence for a glassy skyrmion phase is found \textit{via} cross-correlation analysis, a method which has recently been applied to the analysis of both X-rays and electron diffraction patterns to retrieve information on the local order and symmetry of colloidal systems \cite{wochner2009x,giulialetter,giuliatheo}. In this phase, we reveal also patches of octagonally distorted skyrmion lattice  distribution.
In the skyrmion phase, by locating the position of each skyrmion and generating an angle map of the hexagonal unit cell they formed, we obtain a direct-space distortion map of the skyrmion lattice. This distortion map evidences the presence of orientation-disordered skyrmion lattice domains present within the single crystalline sample. Each domain boundary coincides with a dislocation formed by a seven-five or a five-eight-five Frenkel type defect. The number of such dislocations decreases with increasing magnetic field, and large single-domain regions are formed. The formation of these mesoscopic domains was also filmed with camera-rate (msec) time-resolution. The presence of differently oriented skyrmion lattice domains was observed in spatially separated regions, or in a same area of the sample but at a different moment in time. The formation of regular superstructures arising from co-existing misoriented skyrmion lattices proposed by \cite{langner2014coupled}  is ruled out by these experiments. The observation of split magnetic Bragg peaks in reciprocal space can be explained in terms of a spatial or temporal integration of a fluctuating skyrmion lattice, highlighting the importance of a direct-space, real-time probe.

\section*{Experimental}
	
	A flat and smooth single-crystalline Cu$_2$OSeO$_3$ plate was thinned to 150~nm by the Focused Ion Beam technique. The sample was prepared as a plate with uniform thickness instead of the wedge shaped sample used in previous studies \cite{harada1992real}. This sample geometry prevents significant positional drift of skyrmions as it is common for wedge-shaped samples, due to thickness variations or the temperature gradient. The smoothness and homogeneity of our sample is corroborated by the thickness map shown in the supplementary Fig.\;\ref{Thickness map}. We capture more than 70,000 skyrmions and span their low temperature phase diagram as a function of the external magnetic field. All images were recorded in the (111) sample plane and the magnetic field was applied along the $\langle 111 \rangle$-direction.

\section*{Results}
\subsection{Real- and reciprocal-space maps of different phases.}	 
\begin{figure}
		\vspace*{-5mm}
			\includegraphics[height=0.8\textheight]{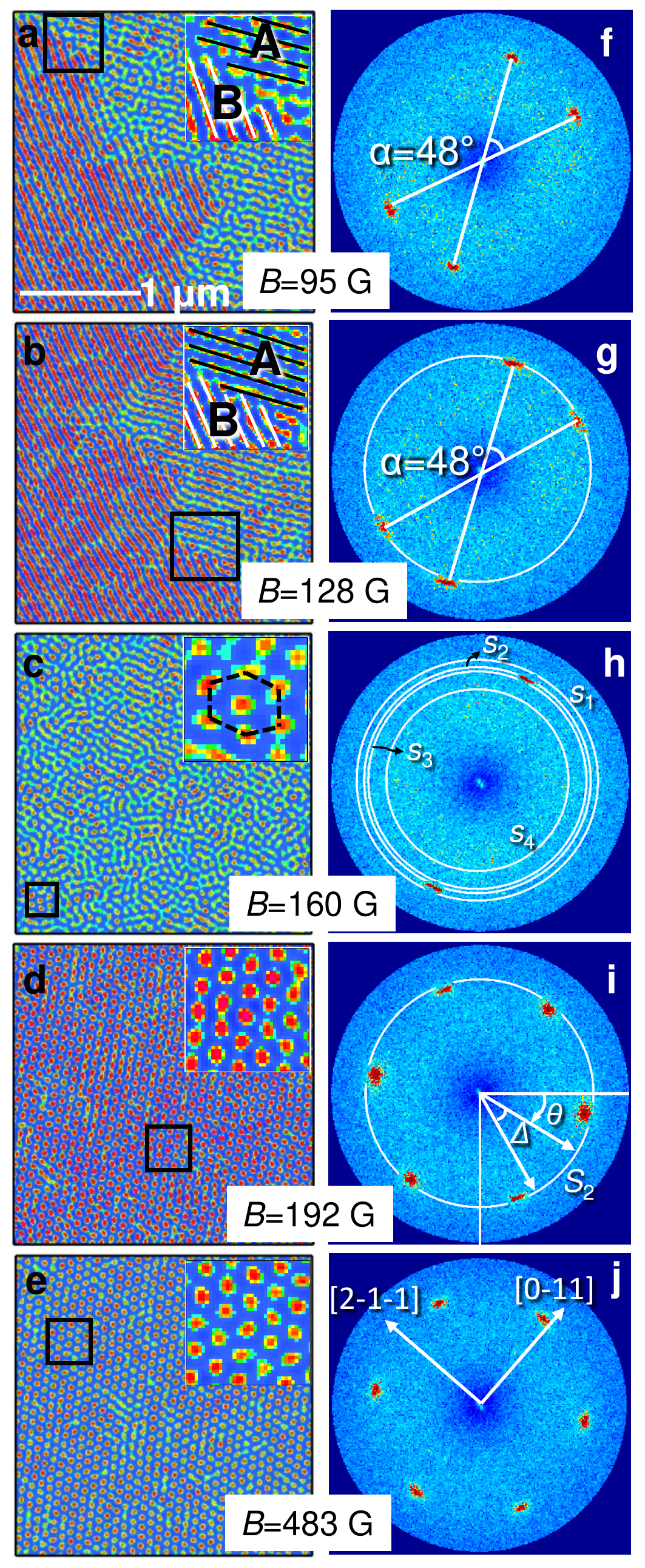}
			\vspace*{-5mm}
			\caption{Magnetic field dependence of the lateral magnetization in a 150~nm thick Cu$_2$OSeO$_3$ single crystal at $\text{T}\approx 7$~K. $2.5\times2.5 $~$\mu$m$^2$ portion of the direct space images (\textbf{a-e}) from the same region at all fields and the  reciprocal space patterns (\textbf{f-j}) obtained from the corresponding $7.3\times7.3$~$\mu$m$^2$ real-space image. The FTs are displayed upto $s = 12\times 10^{-3}\text{\AA}^{-1}$ for clarity. At $B=95$~G and $B=128$~G, the helical phase with two different helical domains are observed. $B=160$~G represents a transition region from the helical to skyrmion phase. At higher fields, a complete skyrmion phase is observed.} 
			\label{Fig1}
		\end{figure}

		Cryo-Lorentz images at a temperature of $\text{T}\approx 7$~K and different magnetic fields are shown in Fig.\;\ref{Fig1} a-e. The images represent a $2.5\times2.5 $~$\mu$m$^2$ zoom of the total $7.3\times7.3$~$\mu$m$^2$ micrograph. The images displayed are treated by a standard Fourier filtering algorithm for better visualization. However, all analyses were performed on the original images.  The raw micrographs of the entire area imaged at all magnetic fields investigated are displayed in the Supplementary Fig.\;S1-7. A further zoom of the real-space image marked by the black solid square in each figure is shown in the insets. The panels f-j depict the reciprocal-space patterns obtained from the corresponding whole and unfiltered real-space image. Every 2D-Fourier Transform (FT) is displayed upto a modulation vector $s = 12\times 10^{-3}\text{\AA}^{-1}$ for clarity. See Supplementary Information for the FT procedure and the images of full 2D-FTs obtained from the corresponding $7.3\times7.3$~$\mu$m$^2$ micrograph.
		
In Cu$_2$OSeO$_3$, the helimagnetic phase develops spontaneously upon cooling below 57~K \cite{PhysRevLett.113.107203,langner2014coupled}, and is visible in Fresnel LTEM as periodically spaced stripes perpendicular to the helices screw axis. At the lowest fields of $B=95$~G (Objective lens off; measured residual magnetic field) and $B=128$~G, we observe two different helimagnetic domains with helices pointing in different directions (Fig.\;\ref{Fig1}a,b). These two domains are marked by A and B in the insets and are characterized by a stripe period of $\approx$~70~nm. Each domain generates a centrosymmetric pair of peaks in reciprocal space, with the propagation vectors for each helical domain rotated with respect to one another by $48^{\circ}$ for both $B=95$~G and 128~G (Fig.\;\ref{Fig1}f,g). The intensity of each pair of centrosymmetric peaks in the FT reflects the degree of occupancy of the corresponding helical domain in the real-space image. These results contrast those of a previous study where only one helimagnetic domain was observed within a probed area of 300\;nm \cite{seki2012observation}. However, periodicity values ranging from 50~nm to 70~nm are reported in the literature \cite{seki2012observation, langner2014coupled, PhysRevB.91.224408}. This variation in the periodicity relates to the strength of the DMI constant D relative to exchange J. It should be independent of sample geometry or size but can depend quite weakly on the applied magnetic field.

At an external field of $B=160$~G, just before the onset of the skyrmion phase, the system is characterized by a glassy distribution with patches of isolated skyrmions and skyrmions in small hexagonal distribution (Fig.\;\ref{Fig1}c and inset). Here, one of the two helimagnetic domains disappears and only one orientation for the helices is found. Accordingly, the corresponding FT shows only one set of centrosymmetric peaks (Fig.\;\ref{Fig1}h).

\begin{figure}
			\includegraphics[width=\linewidth]{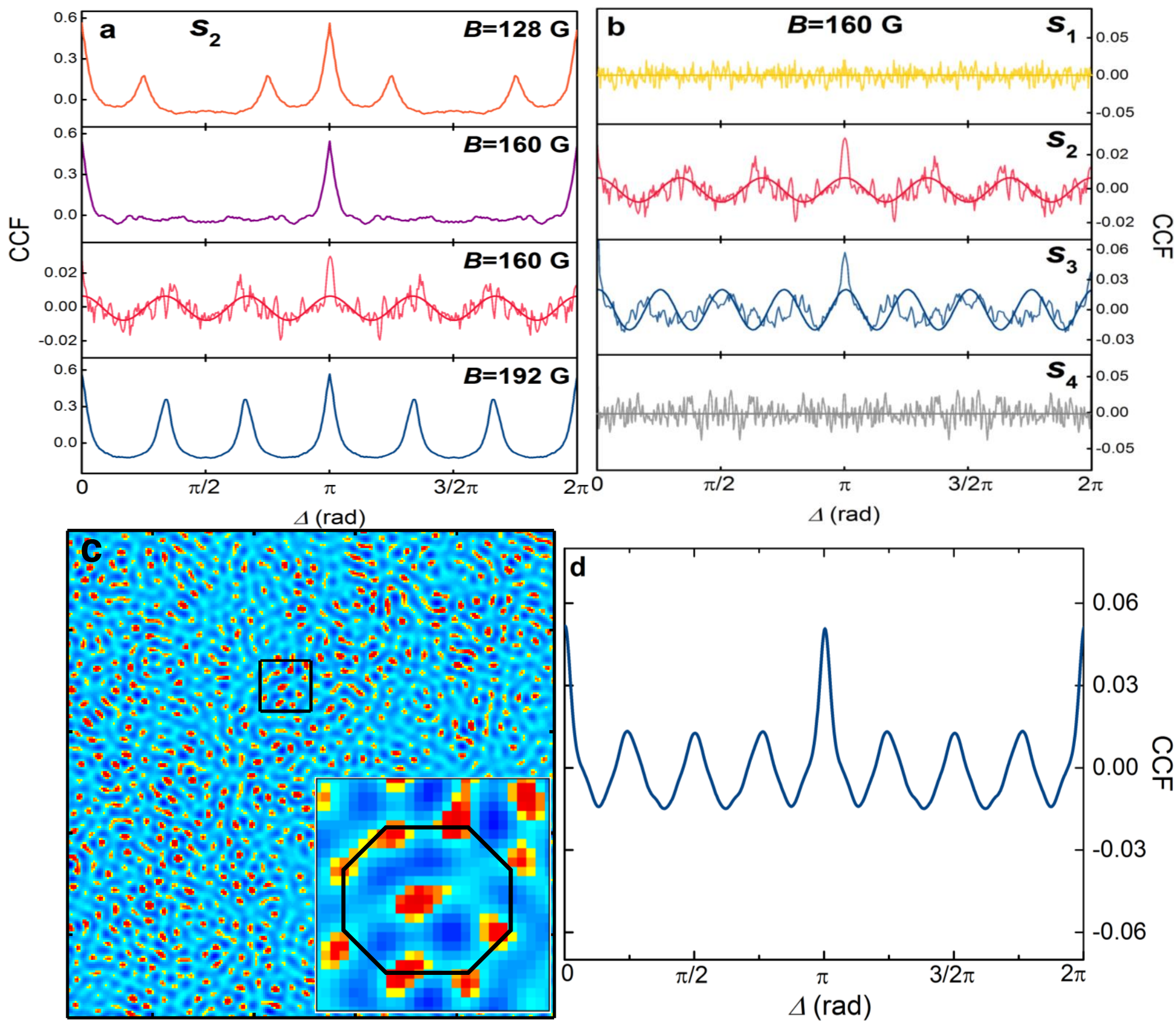}
			\caption{Cross-correlation analysis. \textbf{(a)} Field dependence of the CCF at the scattering vector $s_{2}$ (see text for details). \textbf{(b)} Positional ordering of the magnetic speckle at $B=160$~G for four different scattering vectors.  At the scattering vectors $s_{2}$ and $s_{3}$, respectively, a six-fold and an eight-fold modulations are observed. \textbf{(c)} Real-space image at $B=160$~G highlighting a region with octagonal symmetry. \textbf{(d)} Average CCF of the FT of all the octagonal regions found in the $7.3\times7.3$~$\mu$m$^2$ real-space image at $B=160$~G.} 
			\label{Fig2}
		\end{figure} 

At larger magnetic field strengths, a complete skyrmion lattice forms (Fig.\;\ref{Fig1}d,e and insets). The corresponding reciprocal space image shows the familiar hexagonal pattern (Fig.\;\ref{Fig1}i,j).

\subsection{Cross-correlation analysis.}	
	To investigate the topology of the magnetic structures at all fields, we analyzed the reciprocal space patterns obtained from the FT of the direct-space images (shown in Fig.\;\ref{Fig1}f-j) by means of the Cross Correlation Function (CCF) defined as:
		\begin{equation}
\begin{centering}
C_{s}(\Delta)= \frac{\langle I(s,\;\theta)I(s,\;\theta + \Delta)\rangle\cooltheta-\langle I(s,\;\theta)\rangle\cooltheta^{2}}{\langle I(s,\;\theta)\rangle\cooltheta^{2}}
\end{centering}
\end{equation}
where, $I(s,\;\theta)$ represents the scattered intensity at defined scattering vector $s$, and azimuthal angle $\theta$, while $\Delta$ is the shift between two azimuthal angles. The angle brackets denote averaging over the  variable $\theta$ \cite{altarelli2010x,wochner2009x,wochner2011fluctuations,kurta2012x,giuliatheo}.

Fig.\;\ref{Fig1}i displays schematically how the computation of the CCF is carried out at a selected scattering vector $s_{2}$. Such a method has been successfully used to obtain information on the ordering properties of dilute amorphous systems \cite{wochner2009x} and dense aggregates \cite{giulialetter}. In the presence of a glassy distribution of skyrmions, the cross-correlation function allows the retrieval of information on the local symmetries of the spatial frequency distributions in the sample. We apply this methodology to the FT at the phase transition field, where a glassy distribution of skyrmions is observed.
In panels g-i of Fig.\;\ref{Fig1}, the reciprocal space scattering distribution is shown for three different applied magnetic fields, and a few significant scattering vectors $s_i$ (with i = 1,2,3,4)  are highlighted by white circles. At $B=128$~G, the scattering features related to the helices, which have a pitch of $d_{2}=70$~nm, are found at $s_{2}= \frac{2\pi}{d
_{2}} = 8.97\times 10^{-3}\text{\AA}^{-1}$. At this scattering vector, the computation of the CCF yields the orange trace in Fig.\;\ref{Fig2}a. Upon increasing the magnetic field until the value $B=160$~G two different periodicities are determined in the CCF at $s_{2}$. The first represented by the purple curve is obtained from the two centrosymmetric peaks originating from the helical distribution at this field. When the peaks from the helical arrangement are masked from the diffraction pattern, the underlying diffuse magnetic scattering in the background at $s_{2}$ is accessible. The CCF of the background at $s_{2}$ shows a hexagonal arrangement (red curves) that can be fitted to a harmonic function. The six-fold periodicity retrieved at $B=160$~G from the magnetic speckle in the background reflects the presence of a disordered hexagonal lattice of skyrmions forming in 
the proximity of the phase transition and co-existing with one of the two helical domains. Thus at this field, the CCF unravels an incipient orientational order of the skyrmion  distribution.  Upon entering the skyrmion phase, $B=192$~G, the orientational order is established and the CCF shows sharp peaks (blue trace). Since all the periodicities reported in Fig.\;\ref{Fig2}a are found at the same distance in reciprocal space, this confirms the equivalent periodicity of both the helical and skyrmion lattice magnetic structures.

The positional order of the incipient skyrmion phase at the transition field ($B=160$~G) can be investigated by looking at the scattering vector dependence of the CCF, Fig.\;\ref{Fig2}b. At $s_1$ ($d_{1}=64$~nm) no periodicities are determined (yellow trace), while at $s_{2}$ and $s_{3}$, corresponding to a real-space distance of $d_{2}=70$~nm, and $d_{3}=71$~nm, six fold (red trace) and eight-fold (blue trace) modulated CCFs are found, respectively. The eight fold symmetry is found at a scattering vector corresponding to a slightly larger lattice constant as expected from packing a larger number of skyrmions within the unit cell. At scattering vectors smaller than $s_{4}$ ($d_{4}=85$~nm), no periodicities are determined. This observation suggests that patches of both orientationally disordered hexagonal and octagonal distribution of skyrmions are found, in which the skyrmion - skyrmion distance is included in the range $70 \pm 1$~nm. The presence of such an octagonal distribution can also be seen directly in the real-space image as highlighted in Fig.\;\ref{Fig2}c. The average CCF computed from the FT of all the octagonal regions found in the $7.3\times7.3$~$\mu$m$^2$ real-space image is shown in Fig.\;\ref{Fig2}d indicating a clear eight-fold periodicity.

\begin{figure*}
			\centering
			\includegraphics[width=\linewidth]{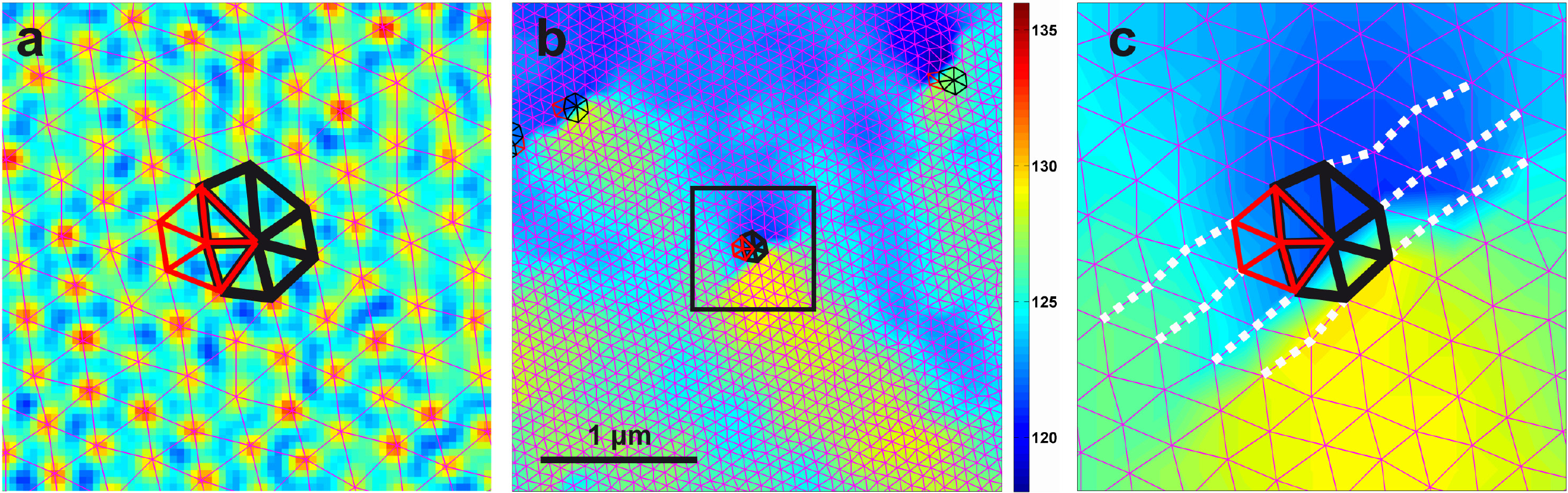}
			\caption{Formation of skyrmion domains. \textbf{(a)} Real skyrmion lattice and a Delaunay triangulated lattice (magenta lines) obtained for $B=483$~G. A skyrmion pair with seven and five neighbors which forms a lattice defect is highlighted with black and red lines, respectively. \textbf{(b)} Spatial angle map of the skyrmion lattice plotted together with the  Delaunay triangulation and defects. \textbf{(c)} A zoom-in of the region marked by square in \textbf{(b)}. The presence of a dislocation line at the domain boundary is evidenced.} 
			\label{Fig3}
		\end{figure*}

\subsection{Formation of skyrmion domains.}
In the skyrmion phase, we evaluate the role of disorder and defects in the lattice by locating the skyrmions in the real-space image and counting the number of nearest neighbors of each skyrmion via Delaunay triangulation. A representative skyrmion lattice at $B=483$~G is shown in Fig.\;\ref{Fig3}a in the background, and the Delaunay triangulated lattice is shown as magenta lines. A perfect skyrmion coordination has a hexagonal symmetry. An imperfect skyrmion coordination can have more or less than six neighbors and forms a lattice defect. A skyrmion pair with seven and five neighbors is highlighted with black and red lines, respectively. A spatial angle map of the orientation of skyrmions (Sections III and IV of the supplementary information) is depicted in Fig.\;\ref{Fig3}b. The formation of a multi-domain skyrmion lattice is readily visible in this map with different colors representing different domains. The Delaunay triangulation and the defects are plotted on the foreground of this map. It is important to note that the domain boundaries coincide with the defects. A zoom-in of a small region marked by a square in panel b is shown in Fig.\;\ref{Fig3}c. The formation of a dislocation at the site of a five-seven defect is evidenced as four lines can be drawn on one side of the seven-five defect, while only three lines can be drawn on the other side. This dislocation line forms the domain boundary between the two orientations of the skyrmion-lattice that are characterized by blue and yellow regions. A similar behavior has been observed at all the magnetic fields and the images are shown in the Supplementary Fig.\;S9-12. However, as the magnetic field is increased the dislocation density decreases, and large single-domain regions form. This highlights the fact that a stronger applied field induces higher levels of order.

\begin{figure*}
			\centering
			\includegraphics[width=\linewidth]{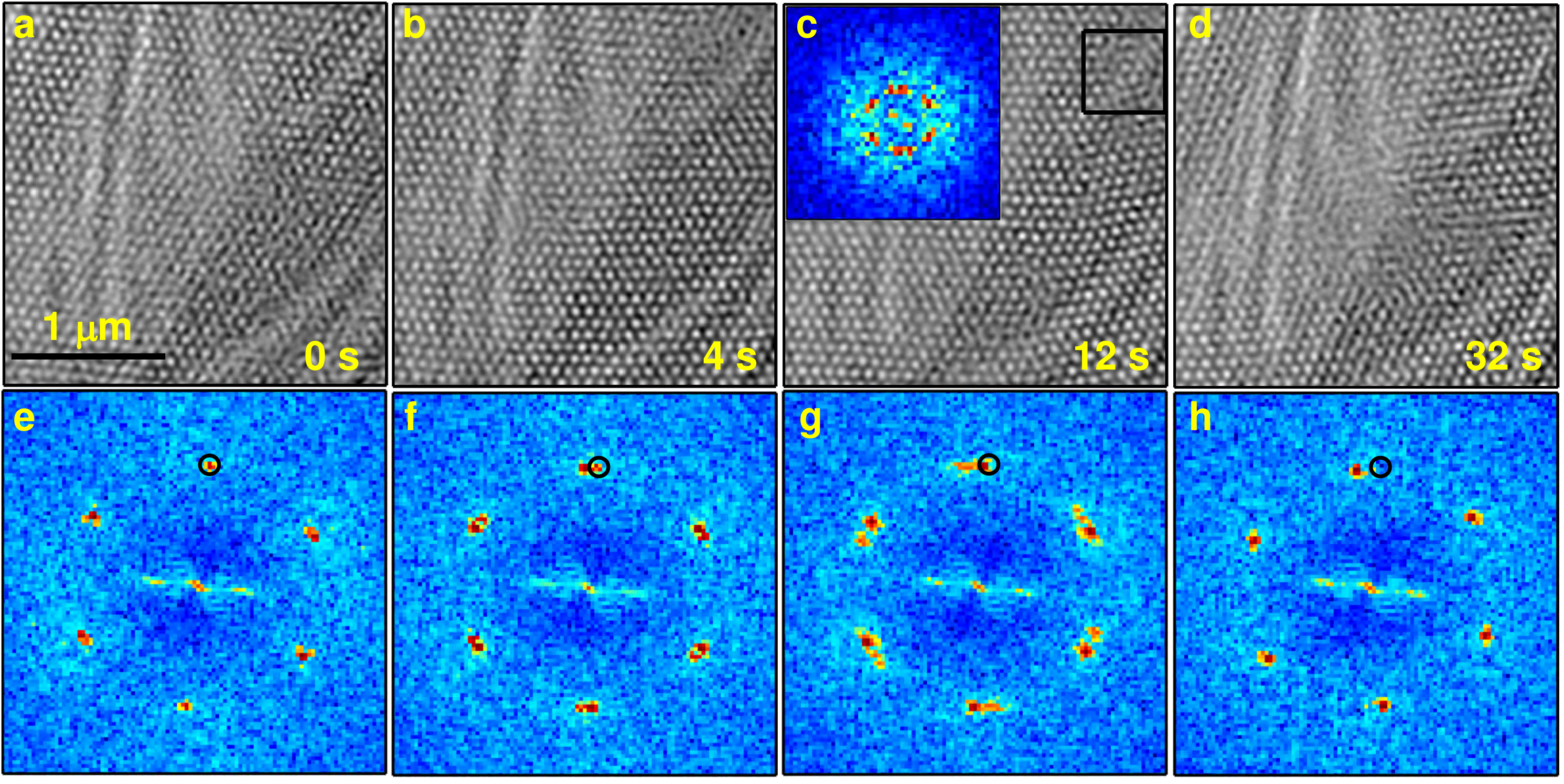}
			\caption{Four different frames of Movie S1 are displayed. Panels \textbf{(a-d)}  depict the real-space images and panels \textbf{(e-h)} represent the corresponding FTs. Fluctuations of the skyrmions lattice and formation of domains with different orientations as a function of time are evidenced by the splitting and unsplitting of the Bragg peaks and their constant change of position.} 
			\label{Fig4}
		\end{figure*}

\subsection{Fluctuation of skyrmion domains.}
To understand the fluctuations of the skyrmion-lattice in real-time, we analyzed a movie acquired for 50\;s. Each frame is exposed for 100\;ms and an image is acquired every 500\;ms. The full movie is shown in the Supplementary  Movie1, four frames at selected time points are displayed in Fig.\;\ref{Fig4}. Panels (a-d) depict the real-space images and panels (e-h) represent the corresponding FTs.
Within this movie, the orientation of the skyrmion lattice is observed to fluctuate and regions of well-distinguished orientations can be separated. At 0\;s (panels a,e), the skyrmion lattice forms a single domain. The corresponding FT shows a relatively sharp hexagonal pattern. The position of a single Bragg peak marked by a black circle can be used as the reference position for marking the deviations in the subsequent frames. At 4\;s (panels b,f) two domains with slightly different orientations are formed. Accordingly, a set of split Bragg peaks is obtained in the reciprocal space map, where the splitting corresponds to the angle between the two domains in the sample with different skyrmion lattice orientations. One of the two sub-peaks is found within the black reference circle indicating that out of the two domains one domain has the same orientation as that found at 0\;s while a second domain with a slightly different orientation has formed within this acquisition time. At 12\;s (panels c,g), the peak separation is larger and also the sub-peaks are further away from the black circle indicating a stronger fluctuation of skyrmion lattice. An interesting point to note in this time frame is that the reciprocal space map of a small region marked by square in panel c also reveals the tendency to have split Bragg peaks (panel c inset). In this particular time and space-point, either the two different orientations are found in the same frame at the same time, or switching between the two has happened within the 100\;ms exposure time. At 32\;s (panels d,h) one of the two domains disappears and forms a new single domain skyrmion lattice which is rotated by about $11^{\circ}$ from the domain formed at 0\;s. This is evidenced by the single Bragg peak found away from the black circle in the FT. Our results clearly emphasize the importance of the time dimension for a proper investigation of the system. Moreover, these observations underline the importance of resolving the skrymion lattice in space and time for revealing its exact topology and the dynamical evolution caused by fluctuations and disordering effects.

\section*{Discussion}
Our results rule out the existence of two superposed skyrmion sub-lattices within a single magnetic structure domain, as proposed in \cite{langner2014coupled}, and imply that the observation of a splitting in the magnetic Bragg diffraction is caused by the spatial and/or temporal overlap of different lattice orientations within the experimental acquisition time.
Remarkably, in specific areas of the sample, we find that disorder can provoke sudden switches between well-defined orientations. This suggests that the energy landscape of the magnetic system has a complex nature with several local minima separated by  subtle barriers. The ability to resolve the switching between these different minima would allow to estimate and control the energy barriers via \textit{ad hoc} external stimuli such as light, electrons or electric fields. Currently, the switching we observed was not entirely resolved due to limitations in the time-resolution of the camera-rate acquisitions. Future experiments with time-resolution in the microseconds to nanoseconds in our ultrafast TEM \cite{Piazza201379, piazza2015simultaneous} should therefore be able to fully resolve this behavior. The application of cryo-LTEM for investigating skyrmion dynamics paves the way to advance towards a promising set of spintronics applications arising from the motion and manipulation of the skyrmion lattice.

\section*{Methods}
High-quality single crystals of Cu$_2$OSeO$_3$ were synthesized by the method of chemical vapor transport redox reactions. The crystal was thinned to 150 nm by the Focused Ion Beam technique. The magnetic structures of the films were investigated by using JEOL JEM-2200FS cryo-LTEM. Images were acquired in the Fresnel mode, \textit{i.e}. defocused imaging \cite{cottet2013quantitative}, so that the objective lens was not utilized for imaging but for applying the magnetic field. The microscope was operated at 200~kV and equipped with a field emission source. The sample was cooled down to $7\--10$~K using the liquid helium TEM holder (Gatan ULTS), and a magnetic field ranging from $95-483$~G was applied normal to the thin plate along the [111] direction. The magnetic field that is parallel to the electron optic axis was directly measured and calibrated at the specimen position.

\section*{Acknowledgements}
We acknowledge Y.~Tokura, F.~Parmigiani, A.~Rosch, C. Reichhardt, C. Olson-Reichhardt  and C.~H\'{e}bert for useful discussions. Work at LUMES was supported by ERC starting grant USED258697 (F.C.), and the NCCR MUST, a research instrument of the Swiss National Science Foundation (SNSF). Work at LQM was supported by ERC project CONQUEST and SNSF (H.M.R.). The work of T.~G. was supported in part by SNSF under division II. The  work of D.~M. was supported by the Scottish Universities Physics Alliance (SUPA).


\pagebreak
\section*{Supplementary Information}

\section{Lorentz micrographs}
Fig.\;S1-7 show the full $7.3\times 7.3 \mu \text{m}^2$ raw Lorentz micrographs imaged at all magnetic fields investigated in this work.

\section{Reciprocal-space images}
The full 2D-FTs at four representative magnetic fields $B=95$~G, 128~G, 160~G and 192~G obtained from the corresponding $7.3\times 7.3 \mu \text{m}^2$ micrograph are displayed in Fig.\;\ref{FigS8}. The FTs are obtained from real-space images which have been processed with Gaussian filtering and windowing procedures. Windowing is carried out to remove the frequency-domain effects that appears due to sharp discontinuities at the original image boundaries. These effects can be removed by elementwise multiplication of the original image by a matrix equal to 1 in the center and trending toward 0 at the edges. With this procedure, the intensity of the image decreases towards a value 0 at the image boundaries.

\section{Skyrmion positions and Delaunay Triangulation}
To improve the contrast between skyrmions and the background, a Gaussian filter is applied to the original Lorentz images. The position of the skyrmions are then located using an algorithm which finds the centroids of the skyrmions. The procedure is thoroughly checked and manually corrected where needed. We count $\approx 9000$ skyrmions and estimate the error in locating skyrmions by the algorithm (due to strongly reduced contrast between the skyrmions and the background in some places) to be less than 10. The Delaunay triangulation (DT) of the skyrmion positions is then calculated using the Delaunay function available in the MATLAB\textsuperscript{\textregistered} library (Fig.\;S9-12). The Delaunay triangulated lattice is plotted on top of the real skyrmion lattice to double-check that the defects found by the DT are real.

\section{Angle map construction}
In the real-space $7.3\times 7.3 \mu \text{m}^2$ image, a windowing function is applied to a small region of $0.73\times 0.73 \mu \text{m}^2$ and a FT of the image is obtained. We then compute the azimuthal dependence of the intensity of the Bragg peaks  determined within a $2\pi$ ring that encompasses the Bragg spots. The procedure is repeated by continuously moving the windowed region horizontally as well as vertically to cover the whole $7.3\times 7.3 \mu \text{m}^2$ micrograph. In the end, a single Bragg peak is selected and its angle is plotted as a spatial map as shown in the main text Fig.\;3 and Fig.\;S9-12.

\section{Thickness map}
The thickness map was obtained by means of Energy Filtered Imaging on the TEM 200kV (Fig.\;S13). Two images, I$_0$ and I$_t$, were acquired using elastically scattered electrons and total scattered electrons, respectively. The thickness map was obtained in terms of the relative thickness $t/\lambda$ via the relation $t/\lambda=\ln(\text{I}_{t}/\text{I}_{0})$: $t$ is the thickness of the specimen and $\lambda$ is the mean free path of the total inelastic scattering \cite{egerton1996electron}. The map was measured at room temperature. The thickest part of the specimen was measured to be 150~nm with scattering  electron microscopy.

\section{Brief description of Supplementary Movie}
\textit{Supplementary Movie1: TimeSequence.avi.} Whole sequence of the skyrmion images shown in Fig.\;4 of the main text as function of time from 0\;s to 50\;s with an interval of 0.5\;s.
\begin{figure}
			\renewcommand\thefigure{S1}
			\centering
			\includegraphics[width=1.1\linewidth]{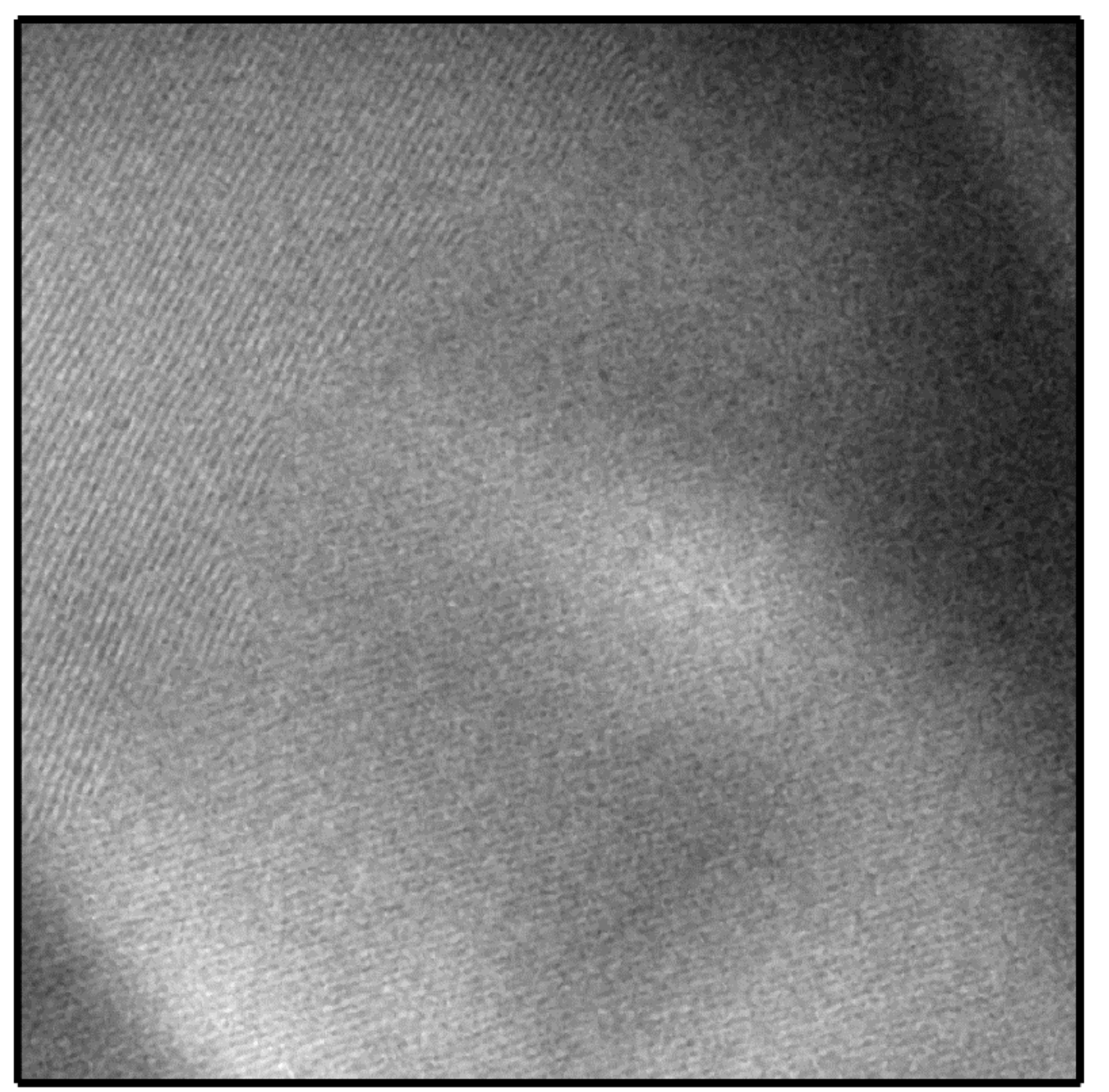}
			\caption{$7.3\times 7.3 \mu \text{m}^2$ real-space image of the helimagnetic phase at $B=95$~G.} 
			\label{FigS1}
		\end{figure}
		
		\begin{figure}
			\renewcommand\thefigure{S2}
			\centering
			\includegraphics[width=1.1\linewidth]{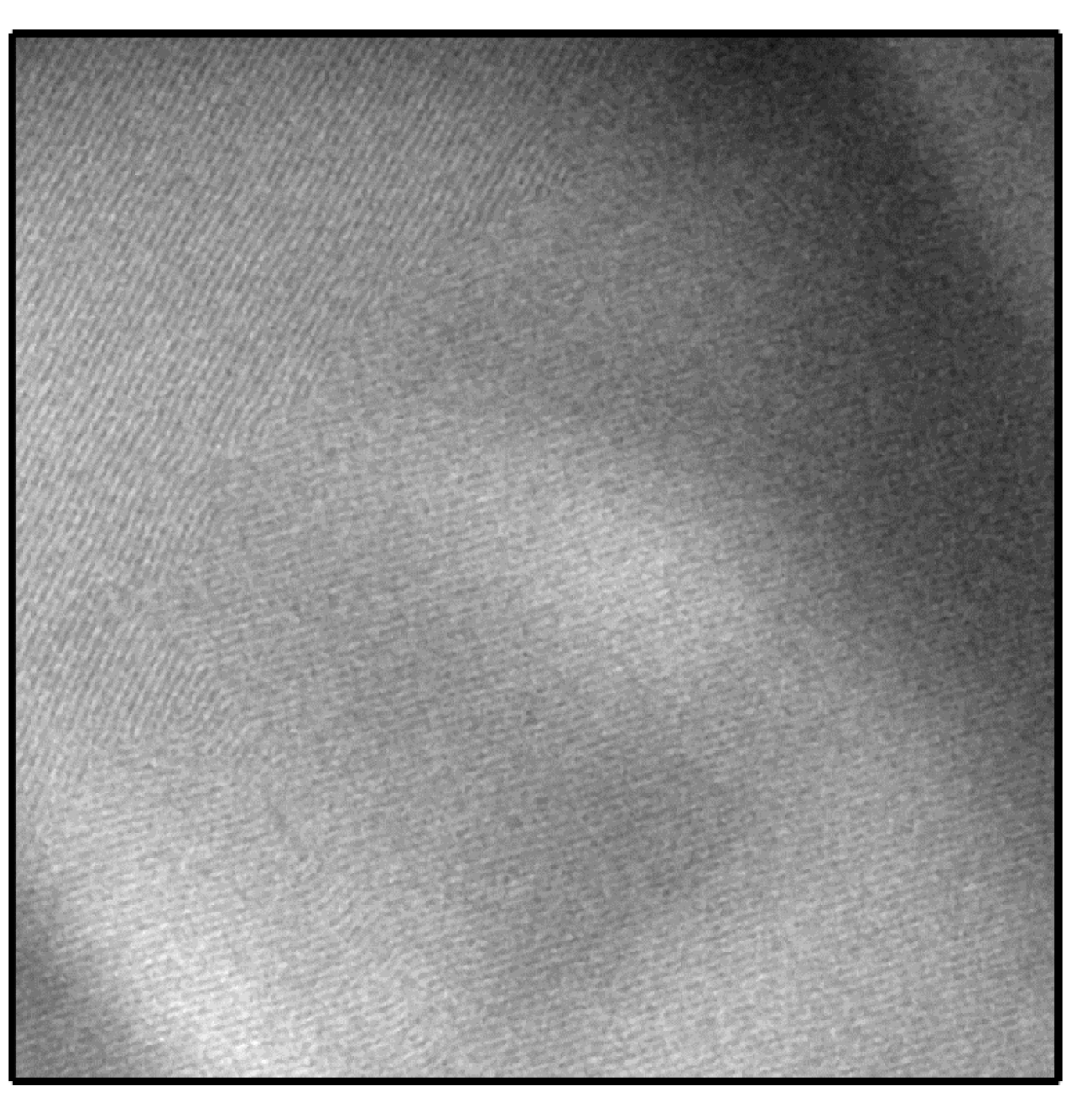}
			\caption{$7.3\times 7.3 \mu \text{m}^2$ real-space image of the helimagnetic phase at $B=128$~G.} 
			\label{FigS2}
		\end{figure}
		
		\begin{figure}[p]
			\renewcommand\thefigure{S3}
			\centering
			\includegraphics[width=1.1\linewidth]{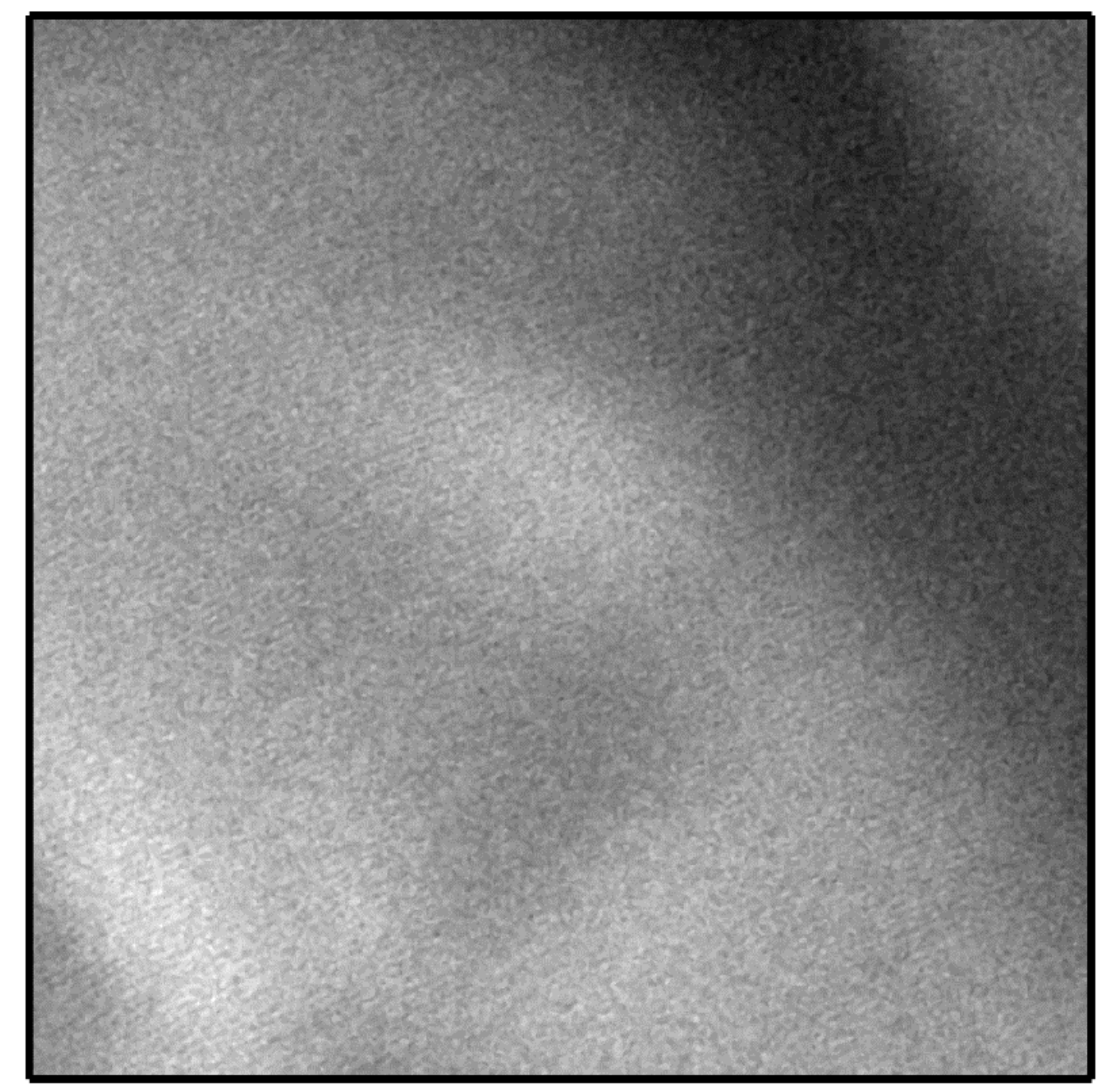}
			\caption{$7.3\times 7.3 \mu \text{m}^2$ real-space image of the glassy phase at $B=160$~G.} 
			\label{FigS3}
		\end{figure}
		
		\begin{figure}[p]
			\renewcommand\thefigure{S4}
			\centering
			\includegraphics[width=1.1\linewidth]{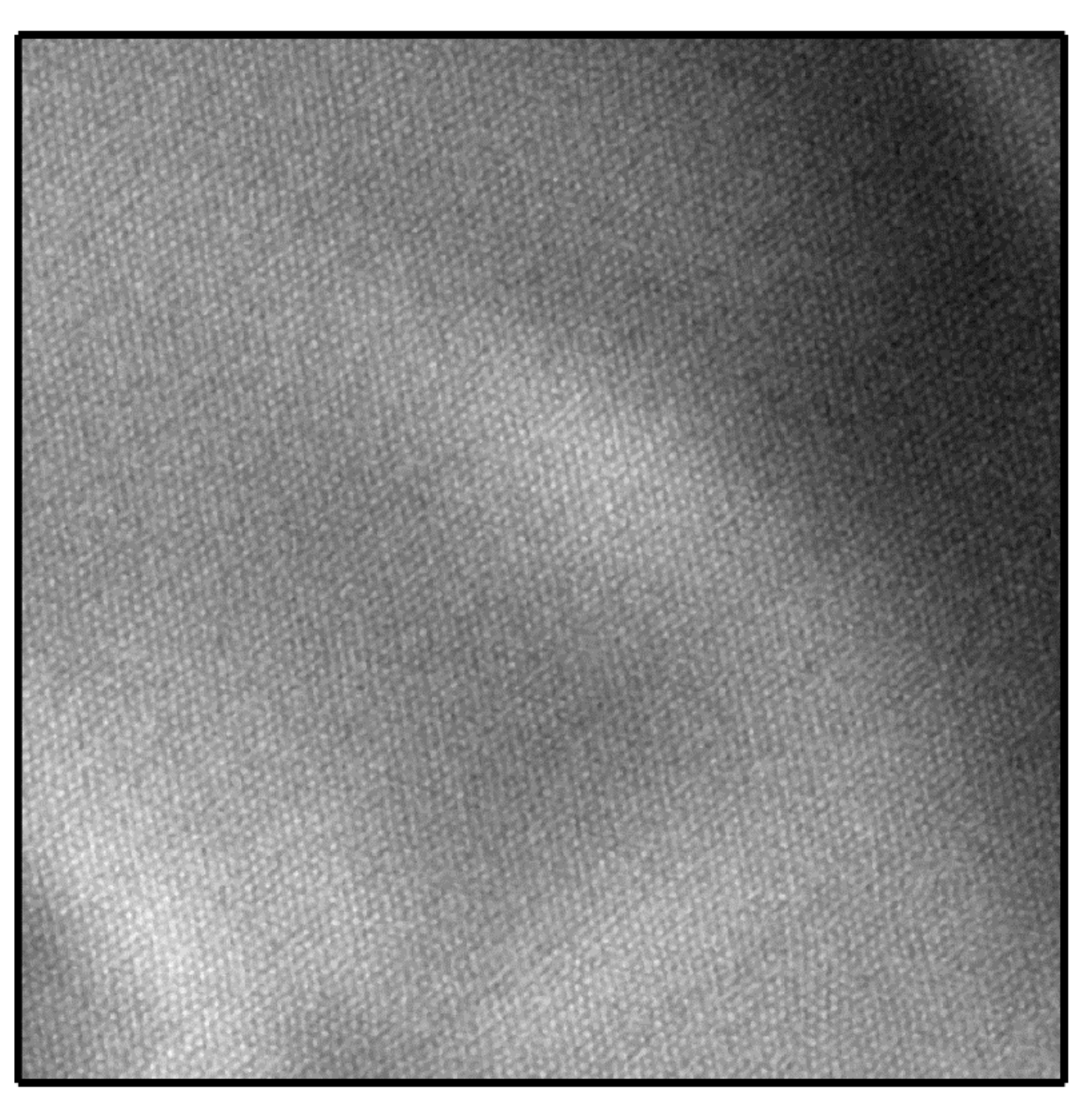}
			\caption{$7.3\times 7.3 \mu \text{m}^2$ real-space image of the skyrmion phase at $B=192$~G.} 
			\label{FigS4}
		\end{figure}
		
		\begin{figure}[p]
			\renewcommand\thefigure{S5}
			\centering
			\includegraphics[width=1.1\linewidth]{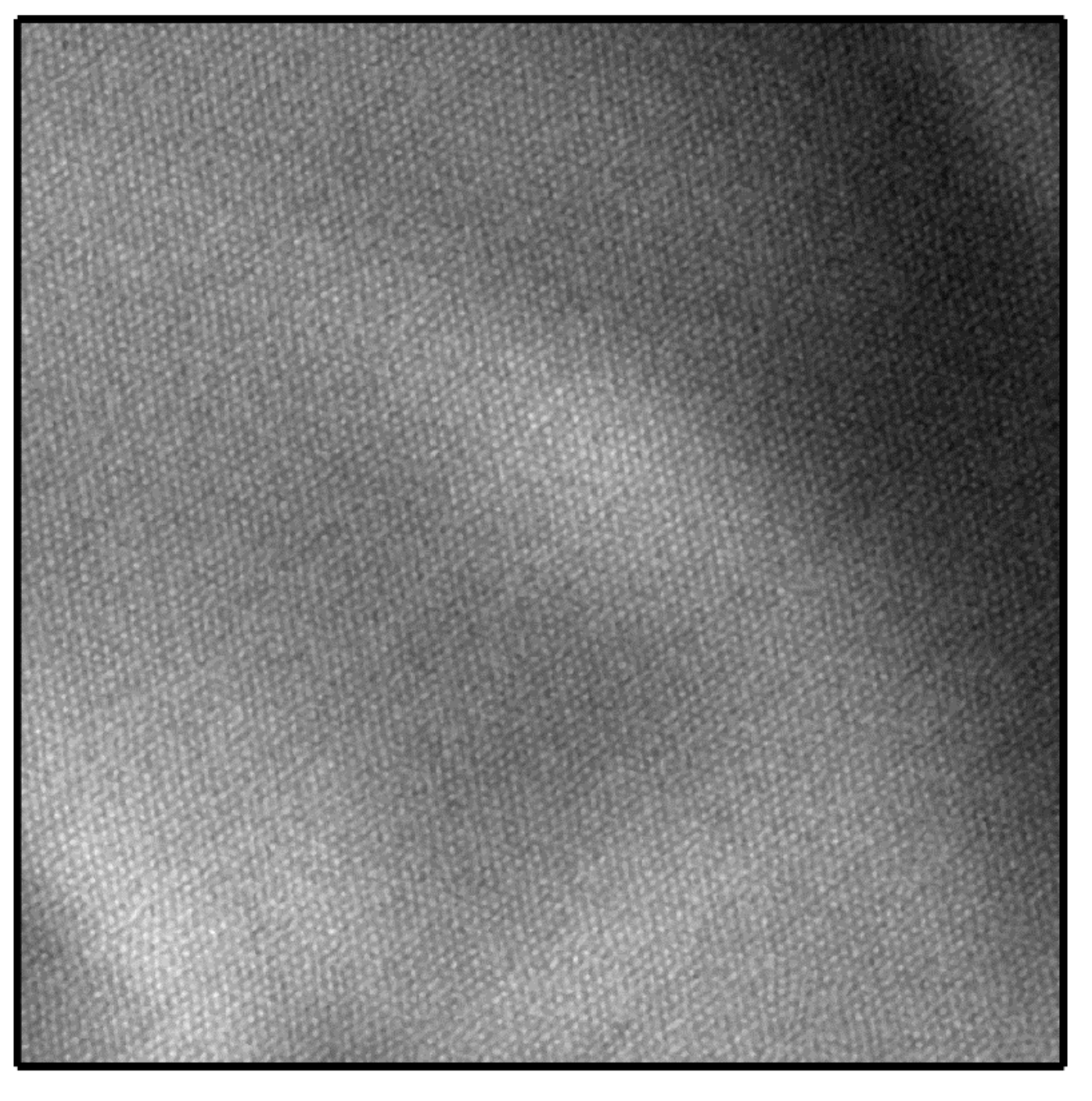}
			\caption{$7.3\times 7.3 \mu \text{m}^2$ real-space image of the skyrmion phase at $B=225$~G.} 
			\label{FigS5}
		\end{figure}

		\begin{figure}[p]
			\renewcommand\thefigure{S6}
			\centering
			\includegraphics[width=1.1\linewidth]{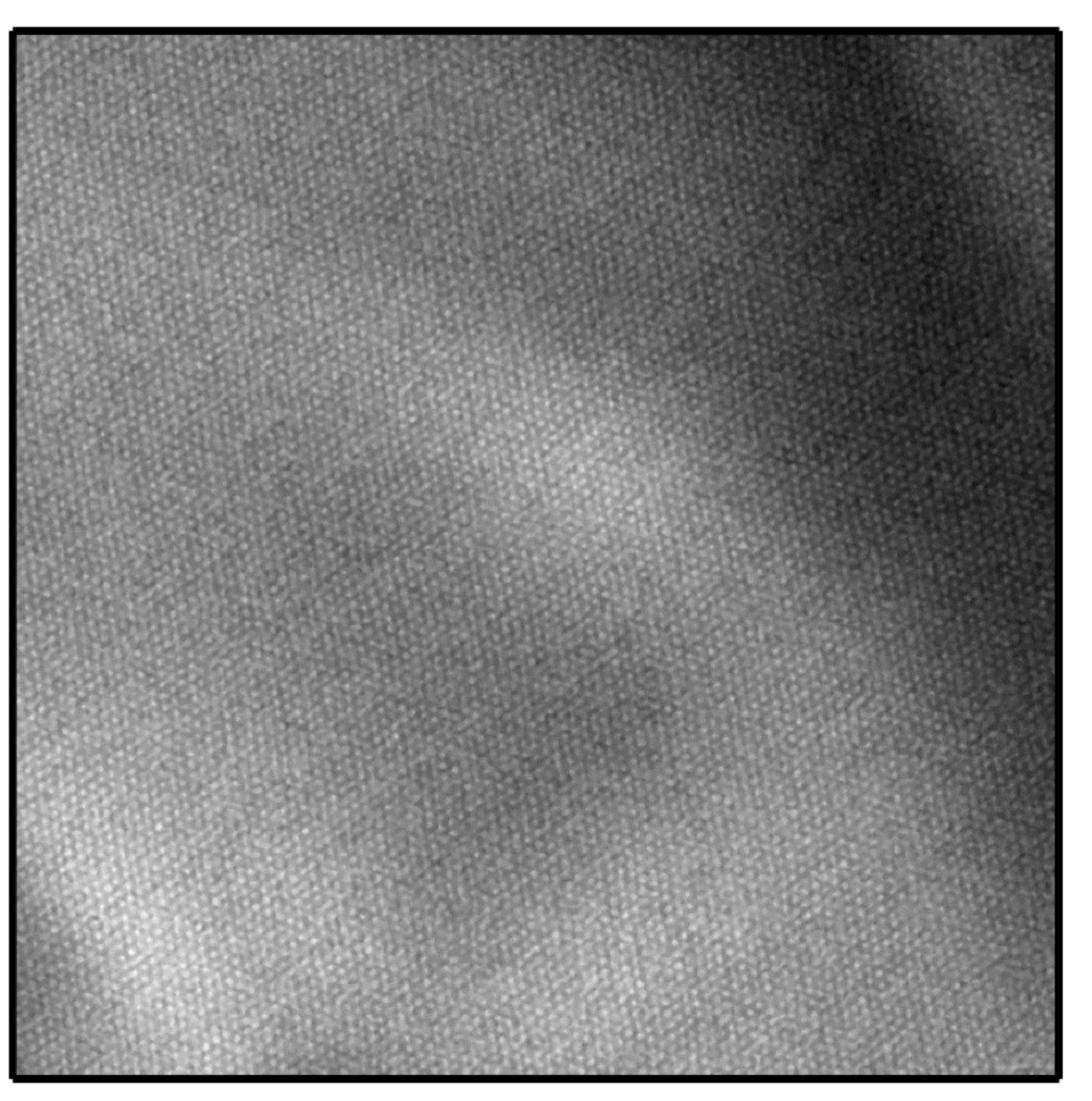}
			\caption{$7.3\times 7.3 \mu \text{m}^2$ real-space image of the skyrmion phase at $B=354$~G.} 
			\label{FigS6}
		\end{figure}

		\begin{figure}[p]
			\renewcommand\thefigure{S7}
			\centering
			\includegraphics[width=1.1\linewidth]{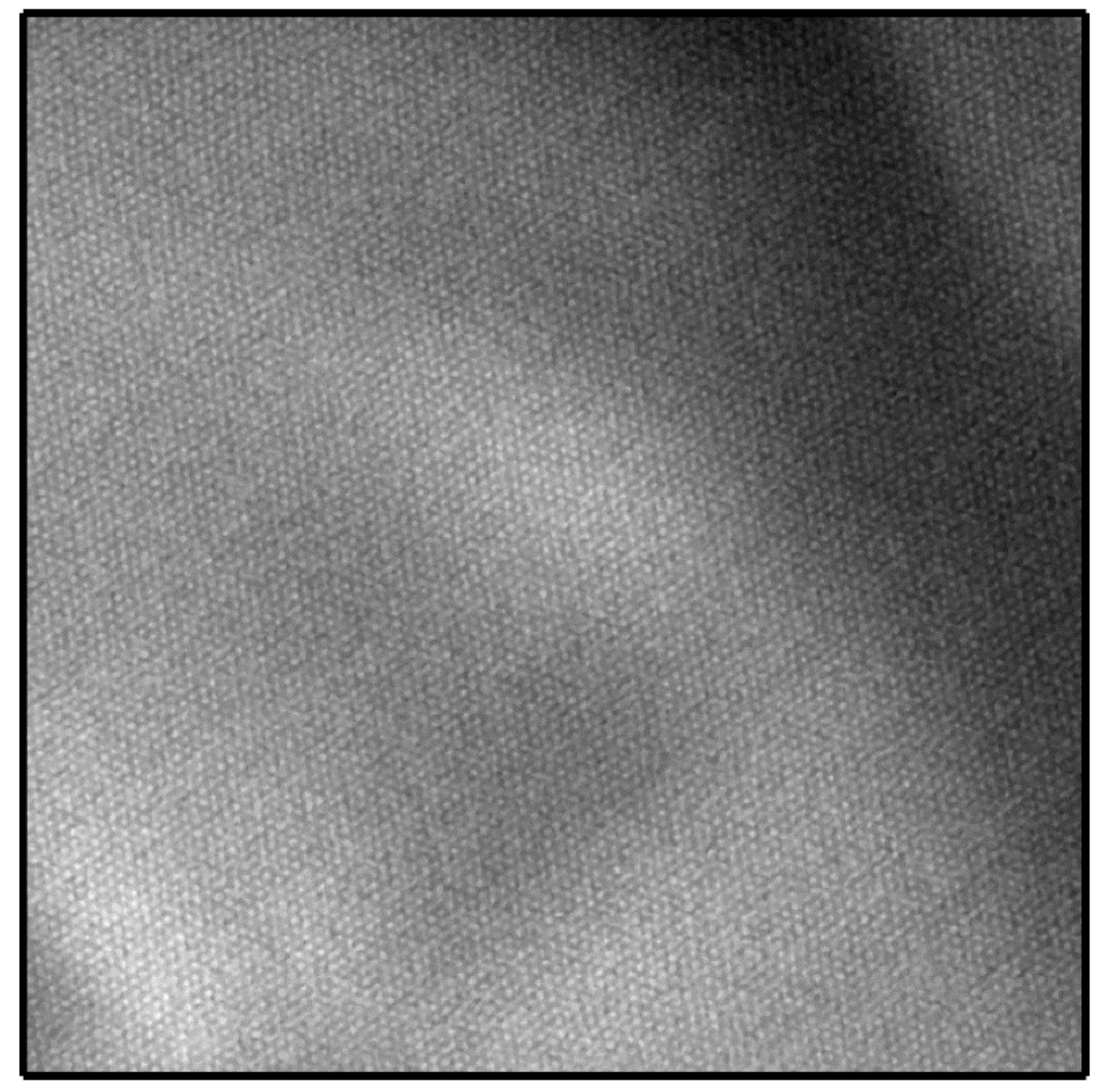}
			\caption{$7.3\times 7.3 \mu \text{m}^2$ real-space image of the skyrmion phase at $B=483$~G.} 
			\label{FigS7}
		\end{figure}
		
		\begin{figure}[p]
			\renewcommand\thefigure{S8}
			\centering
			\includegraphics[width=1.1\linewidth]{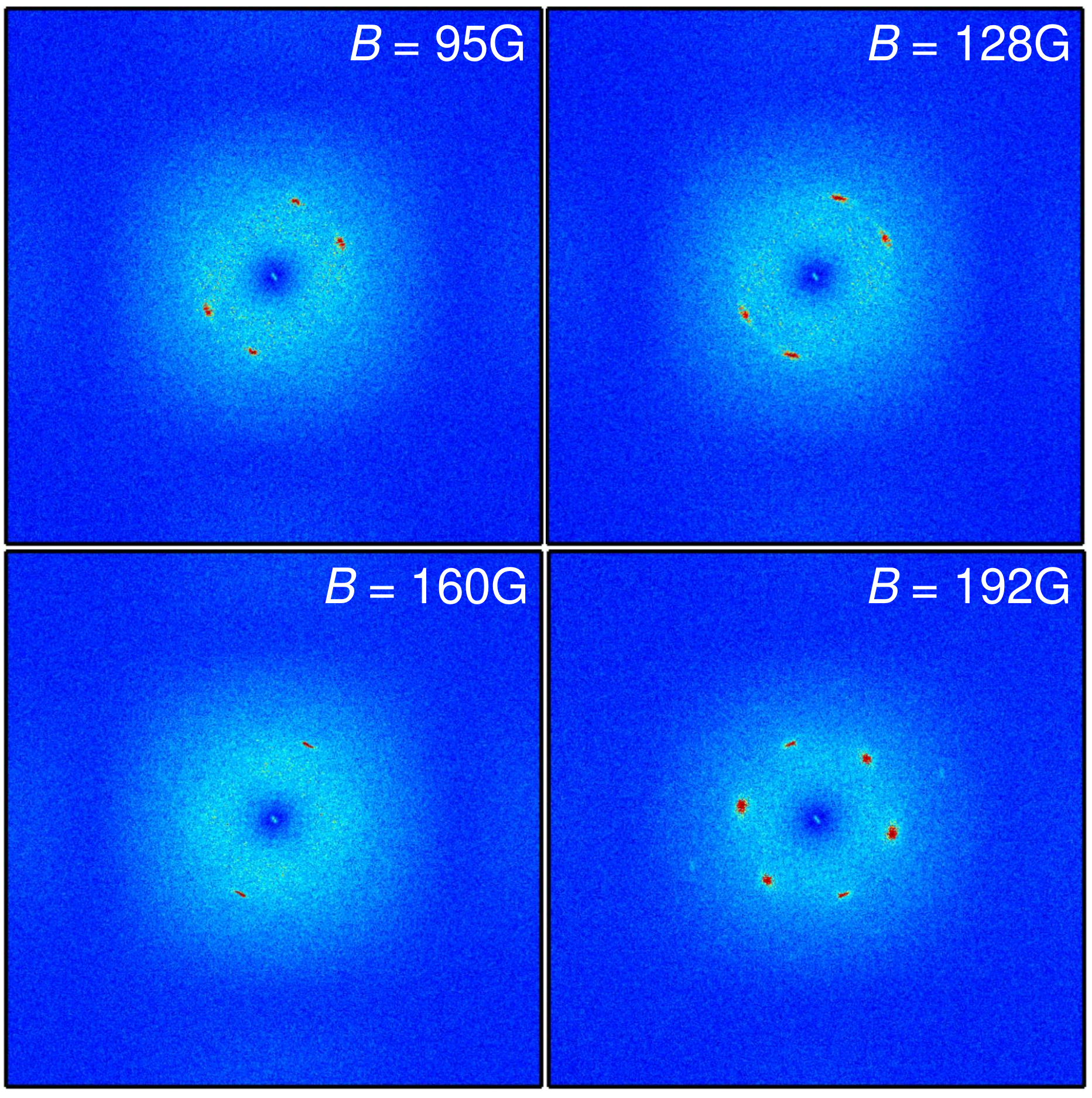}
			\caption{The full 2D-FT's at four magnetic fields $B=95$~G, 128~G, 160~G and 192~G obtained from the corresponding whole real-space image.} 
			\label{FigS8}
		\end{figure}
		
		\begin{figure}[p]
			\renewcommand\thefigure{S9}
			\centering
			\includegraphics[width=1.1\linewidth]{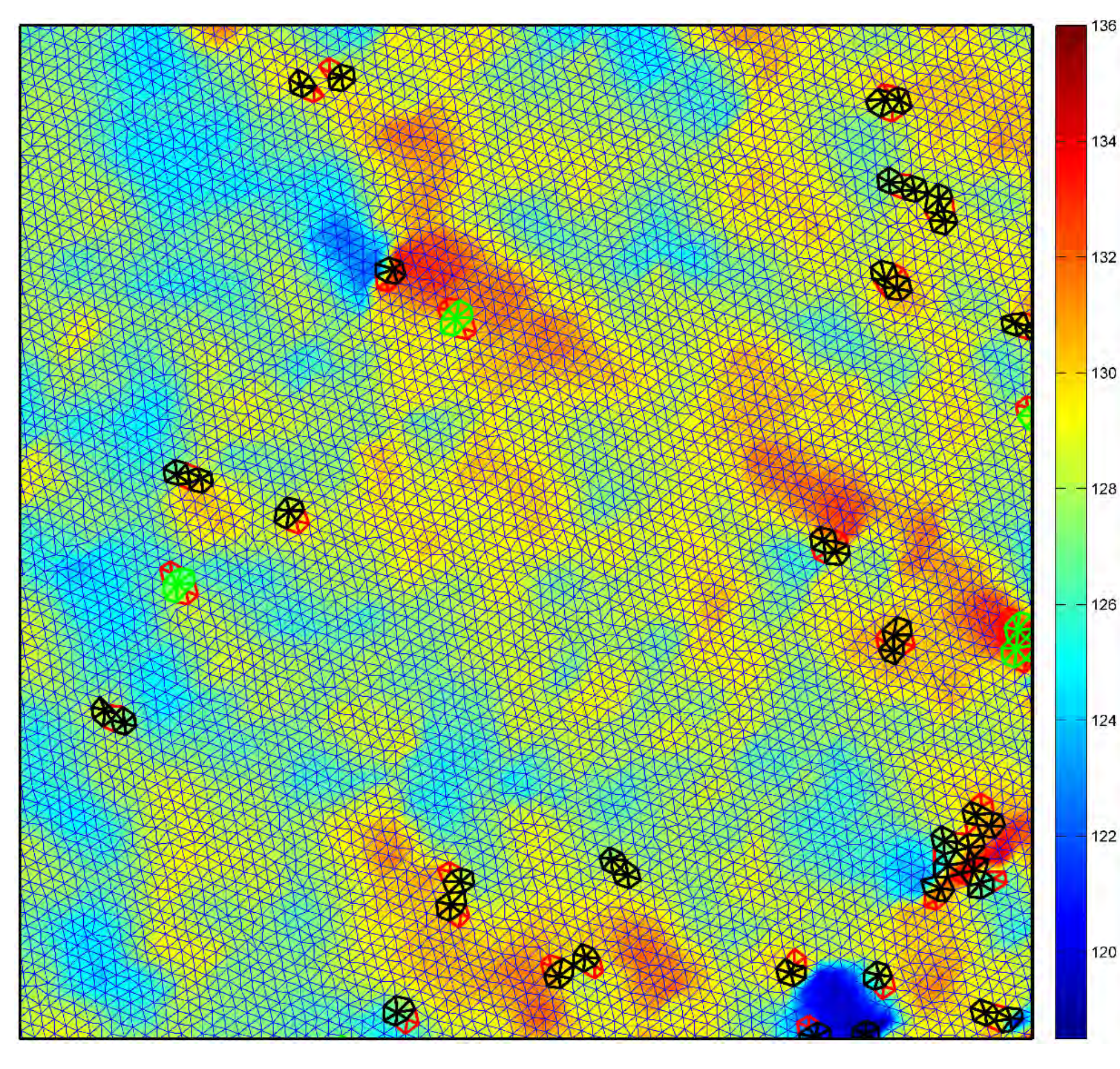}
			\caption{Angle map of the skyrmion lattice at $B=192$~G computed for  the whole $7.3\times 7.3 \mu \text{m}^2$ micrograph plotted together with the Delaunay map and defects.} 
			\label{FigS9}
		\end{figure}
		
		\begin{figure}[p]
			\renewcommand\thefigure{S10}
			\centering
			\includegraphics[width=1.1\linewidth]{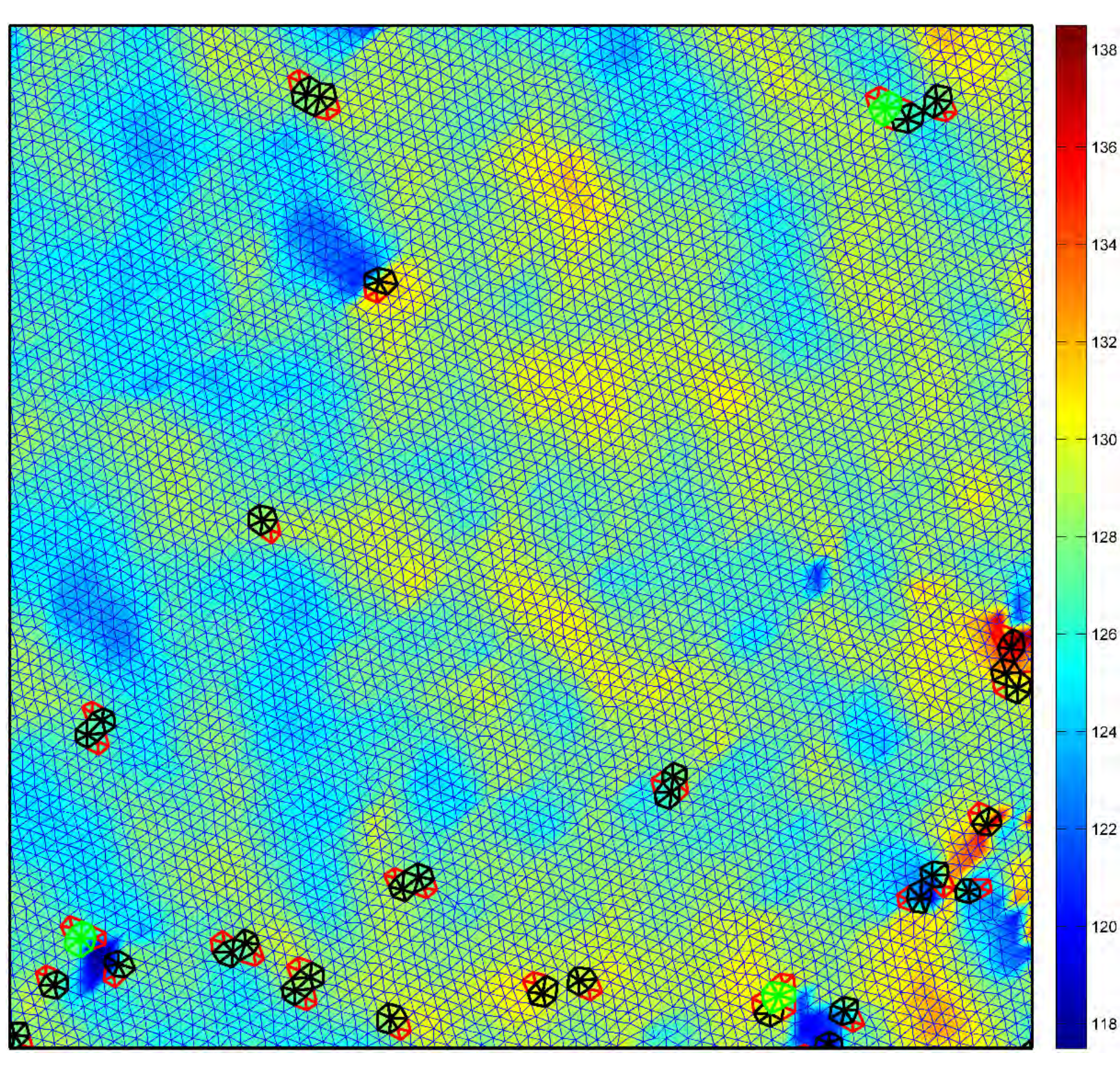}
			\caption{Angle map of the skyrmion lattice at $B=225$~G computed for  the whole $7.3\times 7.3 \mu \text{m}^2$ micrograph plotted together with the Delaunay map and defects.} 
			\label{FigS10}
		\end{figure}
			
		\begin{figure}[p]
			\renewcommand\thefigure{S11}
			\centering
			\includegraphics[width=1.1\linewidth]{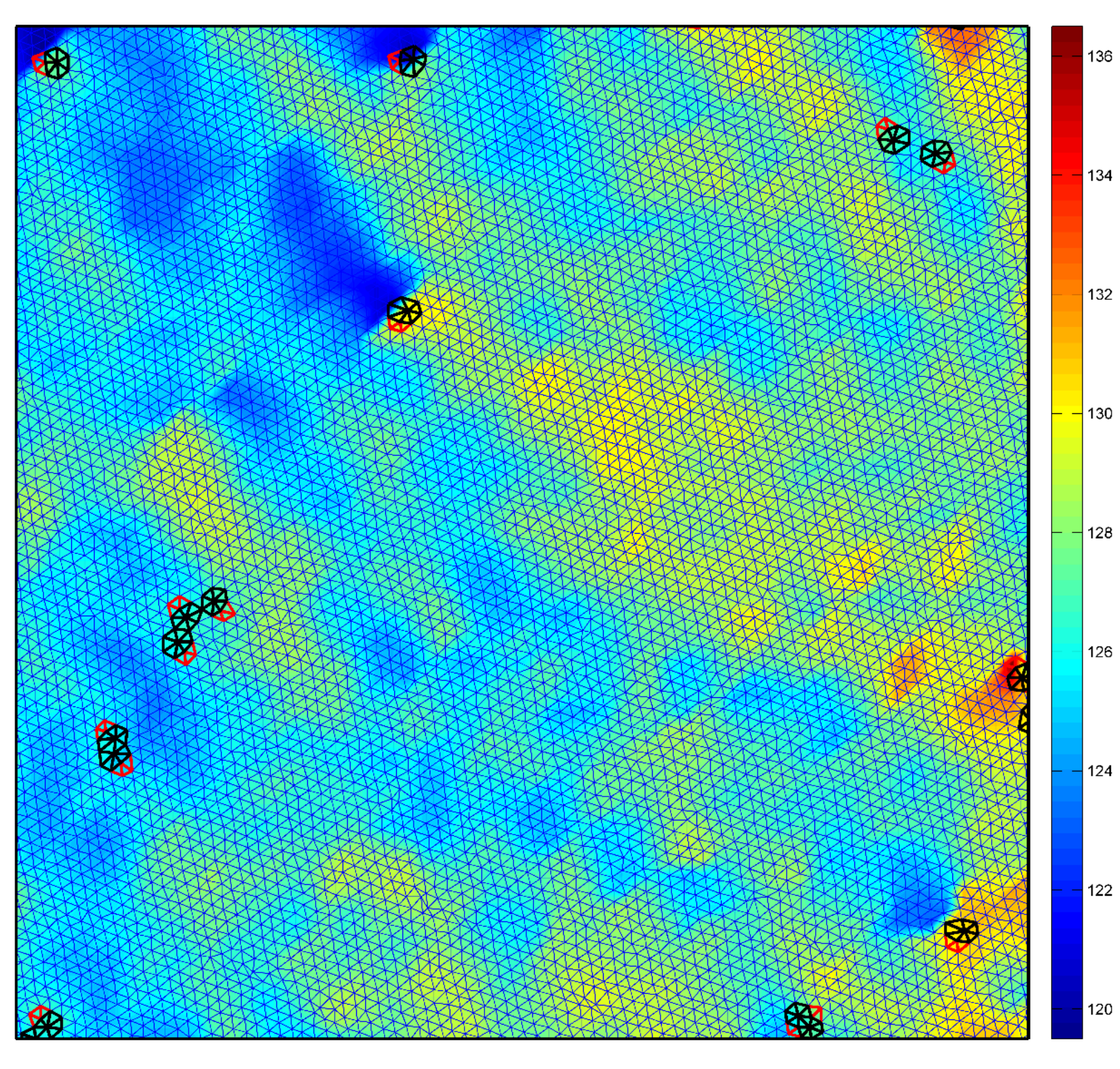}
			\caption{Angle map of the skyrmion lattice at $B=354$~G computed for  the whole $7.3\times 7.3 \mu \text{m}^2$ micrograph plotted together with the Delaunay map and defects.} 
			\label{FigS11}
		\end{figure}
		
		\begin{figure}[p]
			\renewcommand\thefigure{S12}
			\centering
			\includegraphics[width=1.1\linewidth]{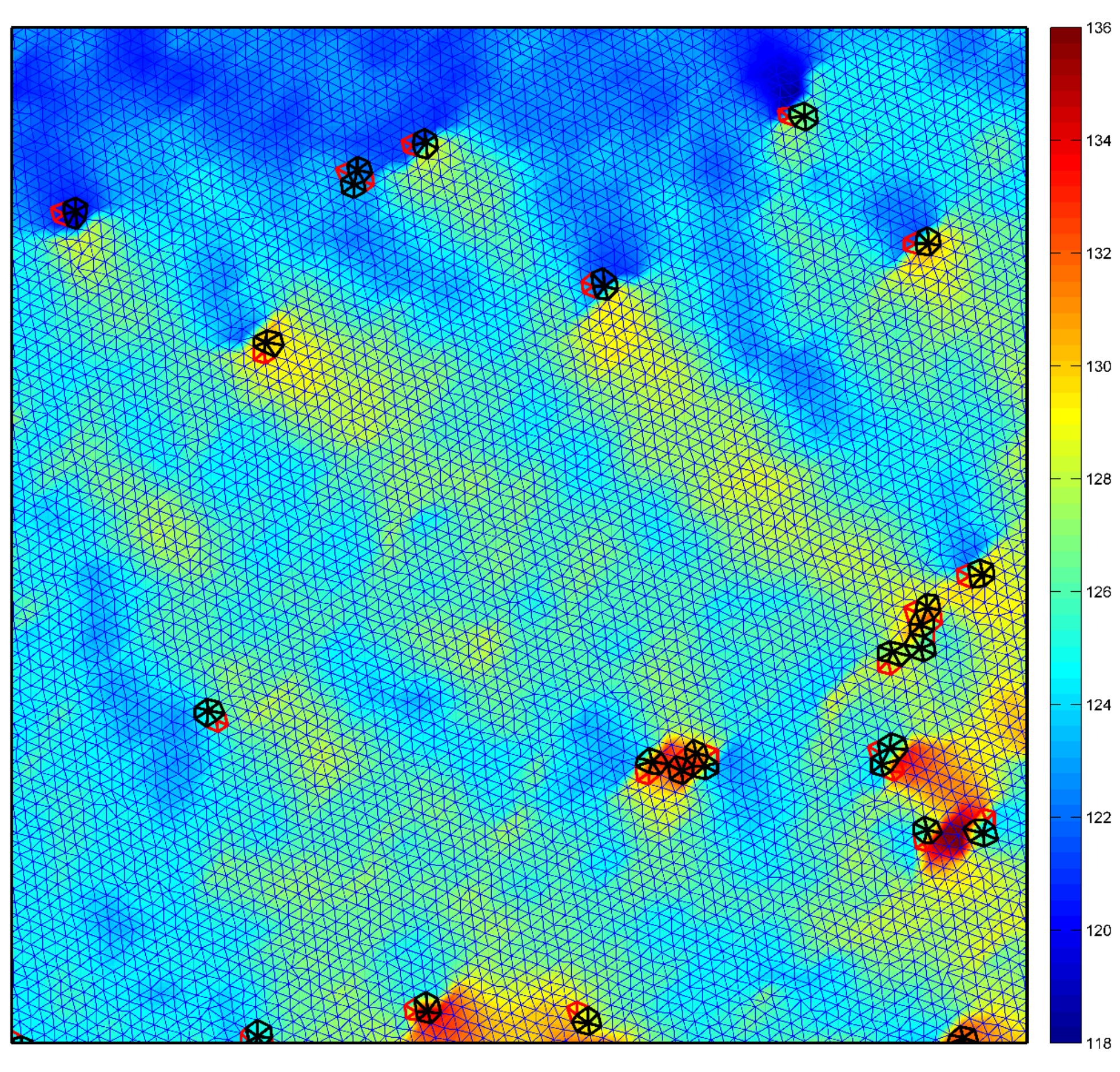}
			\caption{Angle map of the skyrmion lattice at $B=483$~G computed for  the whole $7.3\times 7.3 \mu \text{m}^2$ micrograph plotted together with the Delaunay map and defects.} 
			\label{FigS12}
		\end{figure}
		
		\begin{figure}[p]
			\renewcommand\thefigure{S13}
			\centering
			\includegraphics[width=1.0\linewidth]{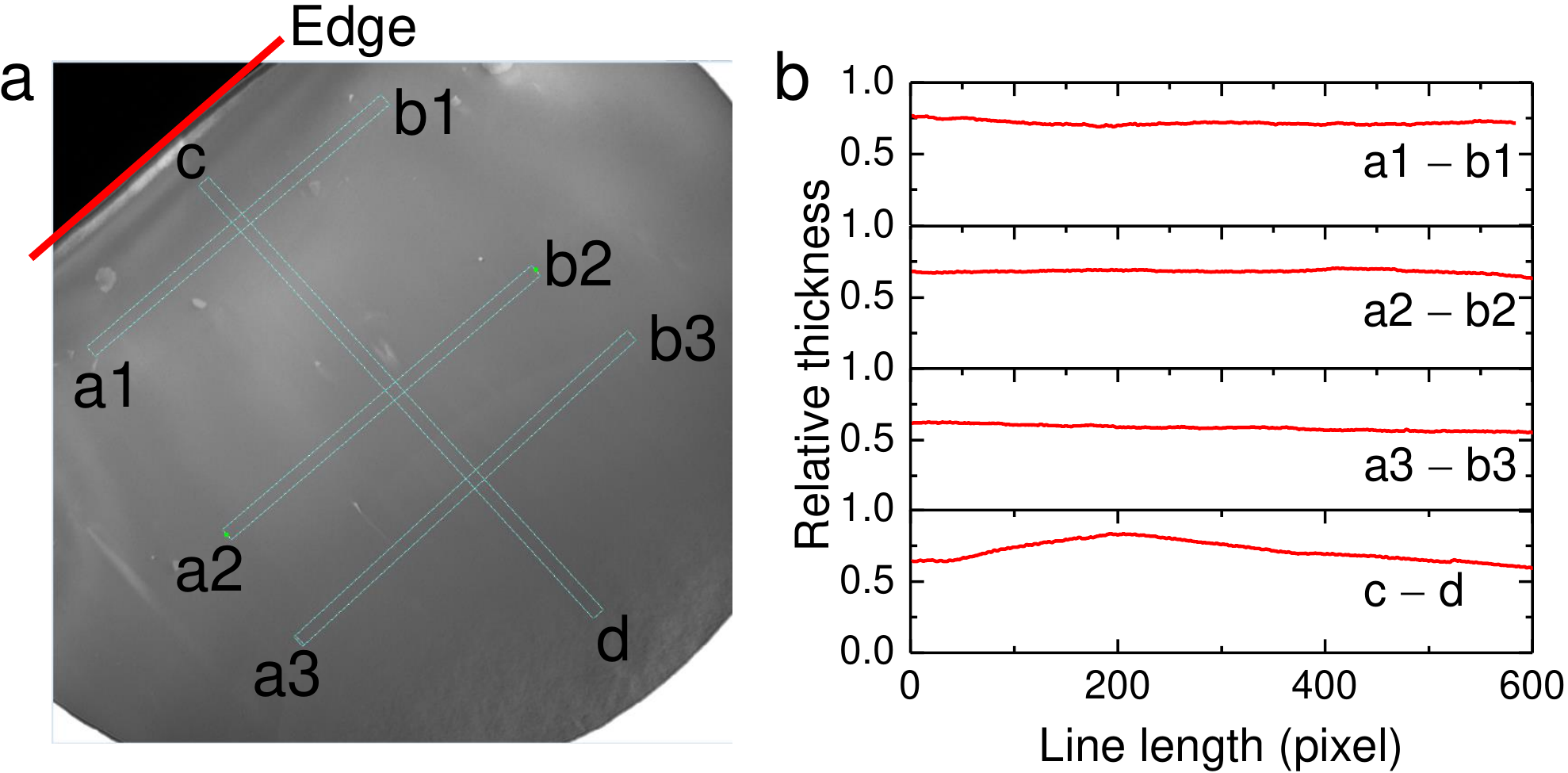}
			\caption{\textbf{(a)} $7.3\times 7.3 \mu \text{m}^2$ thickness map of our sample. Four lines each of 20 pixels thick are marked in the thickness map along which line profiles were recorded. The line profiles are shown in \textbf{(b)}. The relative thickness remains practically constant at 0.6 along the lines parallel to the edge whereas for line perpendicular to the edge, the relative thickness varies between $0.6-0.7$.} 
			\label{Thickness map}
		\end{figure}

\end{document}